\documentclass[aps,pre,twocolumn,longbibliography,amsfonts,amssymb,amsmath]{revtex4-1}
\usepackage{graphicx}
\usepackage{xcolor,soul}

\newcommand{\pd}[2]{\frac{\partial #1}{\partial #2}}
\newcommand{\IInt}[3]{\int_{#2}^{#3}\text{d}#1\;}
\renewcommand{\vec}[1]{\mathbf{#1}}

\newcommand{\gam}{\gamma}

\newcommand{\x}{\vec r}
\newcommand{\X}{\vec R}
\newcommand{\ta}{\tau_\alpha}
\newcommand{\mean}[1]{\langle #1\rangle}
\newcommand{\tobs}{t_\text{obs}}

\begin{document}

\title{Dynamical coexistence in moderately polydisperse hard-sphere glasses}

\author{Matteo Campo}
\author{Thomas Speck}
\affiliation{Institut f\"ur Physik, Johannes Gutenberg-Universit\"at Mainz, Staudingerweg 7-9, 55128 Mainz, Germany}

\begin{abstract}
  We perform extensive numerical simulations of a paradigmatic model glass former, the hard-sphere fluid with 10\% polydispersity. We sample from the ensemble of trajectories with fixed observation time, whereby single trajectories are generated by event-driven molecular dynamics. We show that these trajectories can be characterized in terms of local structure, and we find a dynamical-structural (active-inactive) phase transition between two dynamical phases: one dominated by liquid-like trajectories with low degree of local order and one dominated by glassy-like trajectories with a high degree of local order. We show that both phases coexist and are separated by a spatiotemporal interface. Sampling exceptionally long trajectories allows to perform a systematic finite-size scaling analysis. We find excellent agreement with Binder's scaling theory for first-order transitions. Interestingly, the coexistence region narrows at higher densities, supporting the idea of a critical point controlling the dynamic arrest.
\end{abstract}

\maketitle


\section{Introduction}

Upon cooling, many liquids bypass crystallization and become a supercooled liquid. While dynamics slows dramatically on further cooling, the microscopic arrangement of molecules seems to remain statistically equivalent to that in the disordered liquid~\cite{debenedetti2001supercooled,ediger12}. So far, this conundrum has defied a comprehensive theoretical description and challenges the often successful notion that structure determines even the dynamic properties of a material. Dynamic arrest is a generic phenomenon not only observed in atomic and molecular liquids but also, \emph{inter alia}, in colloidal suspensions~\cite{pusey1986phase, hunter2012physics}, gels~\cite{panyukov1996statistical,scherer1988aging,richard2018coupling}, and polymer melts~\cite{barrat2010molecular,binder2003glass}. There is now a wealth of theoretical approaches~\cite{debenedetti2001supercooled,dyre06,royall2015role,royall18}, which are currently dominated by the idea of a complex free energy landscape characterized by a static length scale over which particles have to move \emph{collectively} in order for the material to relax~\cite{adam1965temperature,kivelson1995thermodynamic,sciortino2005potential,lubchenko2007theory,boue2009predictive}.

However, this dominant role of a static length scale has been challenged recently~\cite{wyart17} through the success of ``swap'' dynamics~\cite{berthier2016equilibrium,ninarello2017models}. Without changing the underlying free energy, it demonstrates that modified dynamic rules are able to substantially postpone the dynamic arrest to lower temperatures and higher densities (but note the rebuttal Ref.~\citenum{berthier19}). In fact, a static length scale is not necessary to rationalize dynamic arrest. Even systems with trivial thermodynamics show glassy behavior as demonstrated explicitly in kinetically constrained models~\cite{garrahan2009first,jack2006space,speck2011space}. Based on this insight, dynamic facilitation theory~\cite{chandler2010dynamics,speck19} provides an alternative explanation in which dynamic arrest is attributed to a disorder-order transition in the ensemble of \emph{trajectories}, much in analogy with the ordering of spins in the Ising model. Over the last 15 years, such dynamical phase transitions have been investigated extensively in lattice~\cite{merolle2005space,garrahan2007dynamical,speck2011space,gutirrez19}, atomistic~\cite{hedges2009dynamic,pitard11,speck2012first,speck2012constrained,turci2017nonequilibrium}, and also quantum~\cite{olmos2012facilitated} model glass formers. However, with the exception of lattice systems~\cite{nemoto2017finite}, the rigorous determination of the nature and order of the transition has received comparably little attention.

The existence of two competing dynamic phases naturally explains the observed strong heterogeneity~\cite{hurley1995kinetic,berthier2011dynamical} in the dynamics of supercooled liquids -- small patches of mobile particles coexist with large patches of immobile particles -- as a pre-transition effect (in analogy with, \emph{e.g.} the wetting of hydrophilic surfaces)~\cite{katira2018solvation}. Moreover, dynamic facilitation theory successfully describes the break-down of the Stokes-Einstein relation~\cite{jung2004excitation}, the asymmetry of heat capacity on cooling and heating~\cite{keys2013calorimetric}, and the scaling between higher-order response functions~\cite{chandler2006lengthscale} measured through dielectric broadband spectroscopy~\cite{albert2016fifth}.

There is now growing evidence~\cite{royall2015role} that supercooled liquids undergo a structural evolution that can be probed through \emph{higher-order} correlations. One route to access these correlations is to identify local structural motifs of $m\leqslant 13$ particles and to measure their population. In several model glass formers, the average population of a certain (system-dependent) motif -- in the following referred to as locally favored structures (LFS) -- shows a strong response upon cooling. Moreover, the population is strongly correlated with dynamics, in the sense that trajectories with a high population of LFS are also ``slow''~\cite{speck2012first}, although a direct correlation of \emph{single particle} displacements with their local structure is much less conclusive~\cite{hocky2014correlation,coslovich2016structure}.

Here we study the arguably most simple glass former, polydisperse hard spheres, which constitutes a paradigmatic model for kinetic arrest driven by excluded volume due to short-ranged repulsive forces. This model plays an important role since exact results in the limit of infinite dimensions are available~\cite{parisi2010mean}, with the hope to gain access to physical dimensions through, \emph{e.g.}, renormalization~\cite{char17,char19}. More importantly, such mean-field approaches typically relate to states deep in the glass, whereas here we are concerned with the supercooled quasi-equilibrium liquid that precedes the dynamic arrest. Moreover, dynamic facilitation aims to treat fluctuations of mobility (and structure) in a non-perturbative fashion, focusing on their large deviations~\cite{touchette2009large}. In the context of trajectory ensembles, hard spheres have been studied in one dimension~\cite{thompson15} and experimentally~\cite{pinchaipat2017experimental}.

Monodisperse hard spheres freeze into a regular face-centered cubic (fcc) crystalline solid with a range of packing fractions for which fluid-solid coexistence is observed~\cite{pusey1986phase}. Crystallization continues for small polydispersity, but with the coexistence region becoming narrower and shifting to higher volume fractions. That crystallization becomes more difficult is easy to understand as spheres with different diameters have to be incorporated into the lattice. For the polydispersity of 10\% studied here, crystallization is already strongly suppressed. Recently, for this polydispersity and packing fractions $\phi\geqslant0.59$ ordering into complex Laves and Frank-Kasper phases in computer simulations has been reported, but requires swap dynamics~\cite{lindquist2018communication,bommineni2019complex}.

To reliably access fluctuations, our numerical approach~\cite{speck2012constrained,turci2017nonequilibrium} is restricted to rather small systems, for which we do not observe ordering into these structures with large unit cells. Nevertheless, it is well-known that the dependence of fluctuations on the system size in the vicinity of phase transitions contains a wealth of information, from which the thermodynamic limit can be extracted~\cite{binder1981finite,binder1984finite,binderlandau}. Here we follow such a strategy and study moderately polydisperse hard spheres as we vary the observation time while keeping the number of particles fixed.


\section{Simulations}

\subsection{Hard spheres}

We consider a system composed of $N$ hard spheres interacting only through volume exclusion. Each sphere $i$ has a different diameter $\sigma_i=0.7+0.3(i-1)/(N-1)$ with $i=1,\dots,N$, leading to an index of polydispersity $s\simeq0.103$, defined as the ratio between the standard deviation and the mean of the diameters. Throughout, we employ dimensionless quantities with length scale $\sigma$ given by the largest diameter and time scale $\sigma\sqrt{m/k_{\text{B}}T}$ with $m$ the mass of the particles. The only control parameter is the packing fraction
\begin{equation}
  \phi = \sum_{i=1}^{N} \frac{\pi\sigma_i^3}{6L^3N}
\end{equation}
defined as the fraction of the total volume $L^3$ occupied by particles. We integrate the equations of motion using Event-Driven Molecular Dynamics (EDMD)~\cite{alder1957phase} in a box of length $L$ using periodic boundary conditions (PBC). We thermostat the system to a temperature $T$ by drawing velocities from the Maxwell-Boltzmann distribution at regular intervals of time $\delta t = 0.1$. The initial configuration is a simple cubic lattice at low density, $\phi = 0.1$, which is then compressed to the final packing fraction $\phi$. We verify equilibration by comparing the compressibility factor of the melt with the prediction from the Carnahan-Starling equation generalized to mixtures of hard spheres~\cite{mansoori1971equilibrium}.

\begin{figure}[t]
  \centering
  \includegraphics{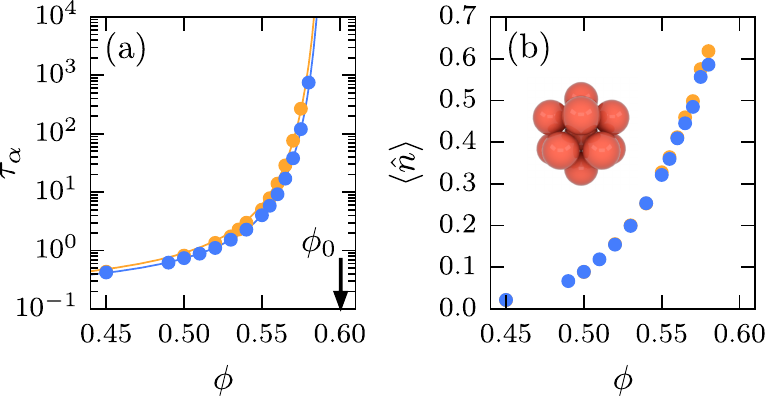}
  \caption{Polydisperse hard spheres. (a)~Structural relaxation times extracted from the decay of the self intermediate scattering function for two system sizes $N=100$ (orange symbols) and $N=1000$ (blue symbols). The colored lines are fits to the VFT expression Eq.~\eqref{eq:vft}, which diverges at $\phi_0$ (arrow). (b)~Average fraction $\mean{\hat n}$ of particles participating in the five-fold symmetric structure 10B shown as an inset.}
  \label{fig:system}
\end{figure}

The structural relaxation is characterized by the relaxation time $\ta$, which is computed from the decay of the self part of the intermediate scattering function 
\begin{equation}
  F(k,t) = \frac{1}{N}\left\langle \sum_{j=1}^{N} e^{i \vec k \cdot[\x_j(t)-\x_j(0)]}\right\rangle,
\end{equation}
where $\x_j(t)$ is the position of particle $j$ at time $t$, and $\vec k$ is the wave vector, the magnitude of which is set to $k\simeq 8$ corresponding to the nearest neighbor shell as obtained from the structure factor. As we increase the packing fraction $\phi$, we observe the typical non-Arrhenius increase of $\ta$ as expected for fragile glass formers, see Fig.~\ref{fig:system}(a). Relaxation times can be fitted to the Vogel-Fulcher-Tammann (VFT) expression
\begin{equation}
  \label{eq:vft}
  \ta(\phi) = \tau_0 \exp\left( \frac{A}{\phi_0 - \phi}\right),
\end{equation}
which diverges at $\phi_0$. We obtain $\phi_0=0.601(1)$, $A=0.195(4)$, $\tau_0=0.119(2)$ for the system with $N=100$ spheres, and $\phi_0=0.601(1)$, $A=0.173(2)$, $\tau_0=0.132(1)$ for $N=1000$. For comparison, the $10\%$ polydisperse hard-sphere liquid has a random close packing of $\phi_{\text{rcp}}\simeq0.67$~\cite{schaertl1994brownian}.

\subsection{Local structure}

Using the Topological Cluster Classification algorithm (TCC)~\cite{malins2013identification}, we count the number of particles that participate in forming defective icosahedra, also termed ``10B'' (see Ref.~\citenum{doye1995effect} for the nomenclature) and report their instantaneous fraction
\begin{equation}   
  \hat{n}(\X) = \frac{1}{N} \sum_{j=1}^{N} h_j(\X), \qquad
  0\leqslant \hat{n}\leqslant 1,
\end{equation}
where $\X\equiv\{\x_j\}_{j=1}^{N}$ is the microscopic configuration of the system. The indicator function $h_j(\X)$ returns 1 if particle $j$ is part of a defective icosahedron, and zero otherwise. The defective icosahedron is a five-fold symmetric structure that was found to be entropically favorable at the local scale in hard-sphere liquids and thus constitutes our LFS~\cite{taffs13,royall2015strong}. Figure~\ref{fig:system}(b) shows the average population $\mean{\hat n}$ of 10B as a function of $\phi$, demonstrating that indeed an increase of local structure is occurring in our system, which strongly correlates with the non-Arrhenius slowing down of the structural relaxation.

\subsection{Sampling of trajectories}
  
We harvest many trajectories of total duration $\tobs$. Trajectories are stored as sequences
\begin{equation}
  x \equiv \left\{ \X(t_k)|t_k = k \Delta t;k=0,\dots,K \right\}
\end{equation}
of $K+1$ configurations sampled at regular intervals $\Delta t$ such that $\tobs=K\Delta t$. We choose the value of $\Delta t$ from analyzing the activated dynamics of single particles, which is characterized by ``jumps'' between the cages formed by their neighbors. We take $\Delta t=5$, the time in which the probability of duration of a jump is reduced by one order of magnitude, see Appendix~\ref{app:jumps} for details.

We characterize each trajectory through the extensive structural order parameter
\begin{equation}
  \mathcal{N}[x] = \sum_{k=0}^{K} \sum_{j=1}^{N} h_j(\X(t_k)),
\end{equation}
the cumulative number of particles participating in the LFS along the trajectory with fraction $n=\mathcal{N}/V$, where $V=N(K+1)$ quantifies the spatiotemporal extent of trajectories. We stress that $n$ is a random variable that changes from trajectory to trajectory and is described by a probability distribution $P(n)$.

\begin{figure}[t]
  \centering
  \includegraphics{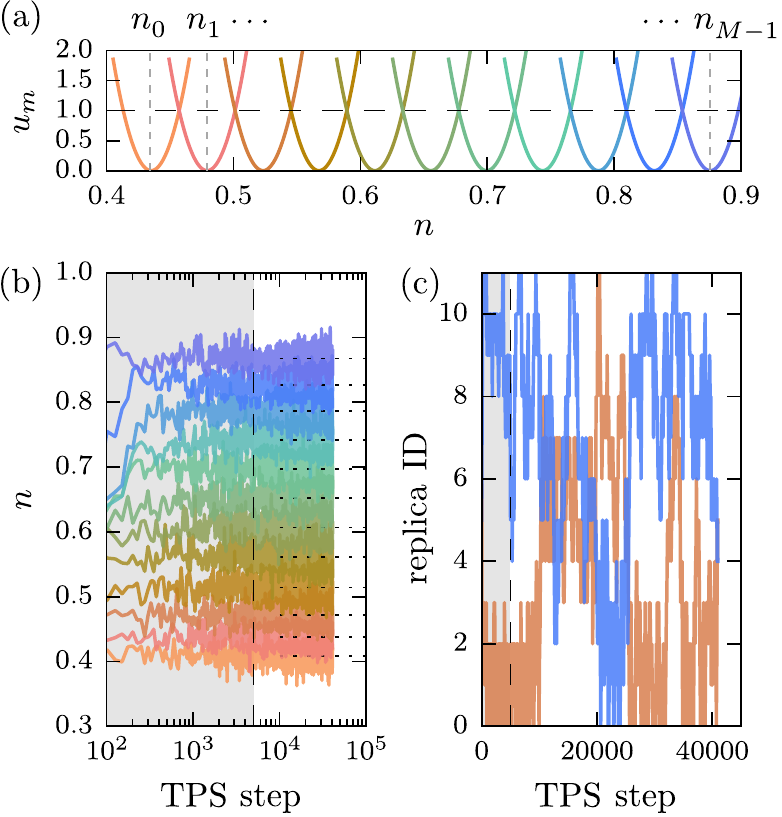}
  \caption{Biased simulations. (a)~Quadratic umbrellas $u_m(n)$ centered around $n_m$. The stiffness $\kappa$ is chosen such that neighboring umbrellas overlap at $u_m=1$ to ensure that exchanges are not too rare. Colors indicate different umbrellas. (b)~Population of 10B over the course of the TPS simulation, for each umbrella. After equilibration, indicated by the grey shaded area, values converge to the interval of $n$ enforced by the umbrella. (c)~Two exemplary replicas, switching between different umbrellas over the course of the simulation.}
  \label{fig:umbrellas}
\end{figure}

\begin{figure*}[t!] 
  \centering
  \includegraphics{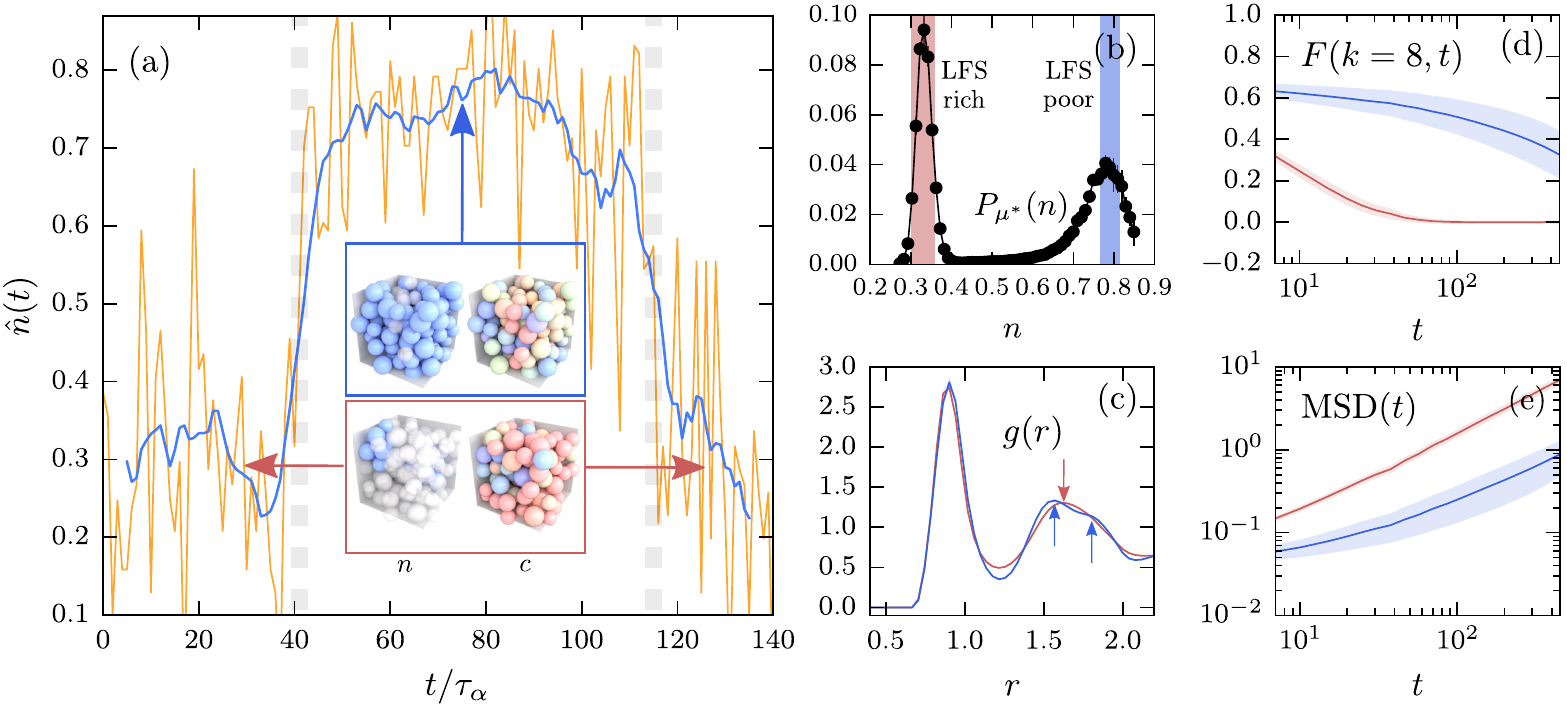}
  \caption{Trajectories have two distinct behaviors. (a)~Instantaneous population $\hat{n}(t)$ of LFS for a single trajectory (orange) at packing fraction $\phi=0.55$, and its running average (blue). The interfaces between the two phases are located at $t/\ta\simeq40$ and $t/\ta\simeq115$. The insets show representative snapshots: left, the blue color indicates whether a particle participates in forming a 10B; right, color indicates the speed of a particle, from slow (blue) to fast (red). (b)~Distribution function $P_{\mu^\ast}(n)$ of the population of 10B at coexistence. Shaded areas indicate the region around the liquid peak (red) and glass peak (blue). (c)~Radial distribution functions obtained from trajectories sampled within the LFS-poor (red) and LFS-rich (blue) region of $P_{\mu^\ast}(n)$. Arrows indicate the positions of peaks. (d)~Intermediate scattering function and (e)~mean-squared displacement for the two phases.}
  \label{fig:coex}
\end{figure*}

Generating trajectories from random initial configurations through EDMD will yield fluctuations of $n$ around the mean $\mean{\hat n}$, which, however, become smaller as the number of particles $N$ and the trajectory length $K$ increases due to the central limit theorem. To access the full distribution $P(n)$, we have to implement importance sampling. To this end, we employ an advanced numerical method based on transition path sampling (TPS)~\cite{bolhuis2002transition, dellago2002transition}, which performs a random walk in the space of trajectories. Starting from an initial trajectory, trial trajectories are generated by applying ``shooting'' and ``shifting'' moves, which are accepted or rejected using the Metropolis criterion. Further details can be found in Refs.~\citenum{speck2012constrained,turci2017nonequilibrium}. Still, for the long trajectories we are interested in one is restricted to rather small system sizes and in the following, all biased simulations will be performed with $N=100$ particles.

In addition, we speed up sampling by preparing $M$ copies of the system, and bias the statistical weight of each copy $m$ using quadratic umbrellas
\begin{equation}
  u_m(n) = \frac{\kappa V^2}{2}(n-n_m)^2,
  \label{eq:umbrellas}
\end{equation}
where the $n_m$ and $\kappa$ are chosen such that the umbrellas cover the interval of $n$ of interest [cf. Fig.~\ref{fig:umbrellas}(a)]. Each copy $m$ explores the region around $\mathcal{N}\sim Vn_m$. Copies are exchanged during the simulation using a replica exchange move, which is again accepted or rejected through the Metropolis criterion. Equilibration can be checked by following the values of representative observables such as $n$ for each replica [Fig.~\ref{fig:umbrellas}(b)] and by checking the number of replica exchanges during the simulation [Fig.~\ref{fig:umbrellas}(c)]. The initial equilibration is discarded for the following analysis.

The full probability distribution $P(n)$ as well as averages are reconstructed using the Multistate Bennett Acceptance Ratio (MBAR) method~\cite{shirts2008statistically,minh2009optimal}, which combines the data from all replicas and removes the sampling bias. In addition, MBAR allows us to calculate the distribution
\begin{equation}
  P_\mu(n) = \frac{1}{Z(\mu)}P(n)e^{\mu Vn}
\end{equation}
of reweighted trajectories with dynamic partition function $Z(\mu)$ as well as expectations $\mean{\cdot}_\mu$. Here, $\mu$ plays the role of an effective chemical potential. Hence, by changing $\mu$ to positive values, we select trajectories with an exceptionally high population of the LFS. This approach has been termed the $\mu$-ensemble~\cite{speck2012first,turci2017nonequilibrium}.


\section{Results}

\subsection{Trajectories exhibit two states distinguished by structure and dynamics}
 
The evolution of the instantaneous population $\hat{n}(t)$ per configuration along a single exemplary trajectory is shown in Fig.~\ref{fig:coex}(a) for packing fraction $\phi=0.55$ and trajectory length $\tobs/\ta\simeq 140$. Within this trajectory, we observe a change at $t/\ta\simeq 40$ from a state with a few structural motives ($\hat n\sim0.3$) to a state in which these are abundant ($\hat n\sim0.75$). In the following, we will identify these structurally distinct states with genuine phases and label them LFS-poor and LFS-rich, respectively. At $t/\ta\simeq 115$ there is a transition back to the initial phase so that the system spends roughly half of the time in both phases. The order parameter of the trajectory is therefore $n\sim0.5$. It corresponds to a local minimum of the probability distribution $P_{\mu^\ast}(n)$, where $\mu^\ast$ is tuned to maximize the fluctuations of $n$ as measured by the susceptibility
\begin{equation}
  \label{eq:chi}
  \chi(\mu) = V[\mean{n^2}_{\mu}-\mean{n}_{\mu}^2] = \pd{\mean{n}_{\mu}}{\mu}.
\end{equation}
Setting $\mu=\mu^\ast$ corresponds to a ``coexistence'' of the dynamic phases in analogy with conventional phase transitions. Correspondingly, the probability distribution $P_{\mu^\ast}(n)$ shown in Fig.~\ref{fig:coex}(b) acquires a bimodal shape.

To better characterize the two phases, we compute the radial distribution function $g(r)$, the intermediate scattering function $F(k,t)$, and the mean-squared displacement $\text{MSD}(t)$ selecting trajectories that contribute to the peaks of $P_{\mu^\ast}(n)$ [shaded areas in Fig.~\ref{fig:coex}(b)]. The radial distribution function $g(r)$ [Fig.~\ref{fig:coex}(c)] shows only minor variation, indicating that no global structural change is occurring in the system. We observe, however, a noticeable splitting of the second peak of $g(r)$, which can be related to the increase of icosahedral ordering~\cite{shen2009icosahedral}. In contrast, the intermediate scattering function [Fig.~\ref{fig:coex}(d)] and the mean-squared displacement [Fig.~\ref{fig:coex}(e)] reveal that dynamics in the LFS-rich phase are much slower, with a relaxation time almost two orders of magnitude larger than for trajectories in the LFS-poor phase. We stress that the density of the system is the same in the two phases, thus the slowdown is directly related to a structural change. Justified by the dramatic difference in the dynamics, we will denote the LFS-poor phase as ``liquid-like'' and the LFS-rich phase as ``glass-like''.

\subsection{The transition is first-order}

\begin{figure*}
  \centering 
  \includegraphics[width=1.0\textwidth]{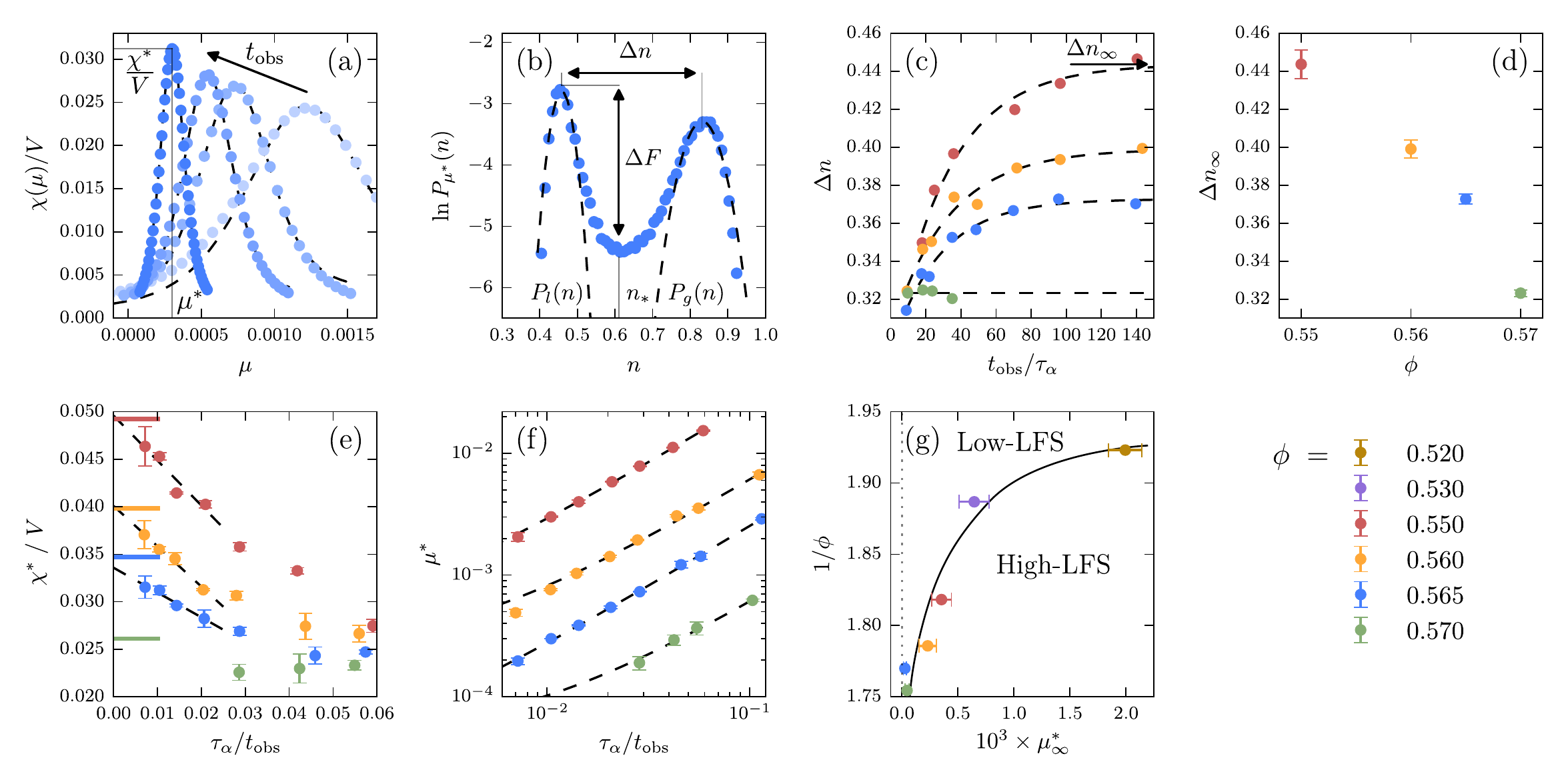}  
  \caption{Finite-size scaling. (a)~Susceptibility $\chi(\mu)$ of the system as a function of the chemical potential $\mu$ for $\tobs / \ta = 22, 35, 48, 96$ (right to left) at $\phi = 0.565$. We fit Eq.~\eqref{eq:chi:fss} (dashed lines) to the numerical data (symbols) with $\Delta n$ as fit parameter and using the bulk susceptibilities $\chi_\text{l}$ and $\chi_\text{g}$ obtained from (b)~fitting Eq.~\eqref{eq:P:gauss} (dashed lines) to the distribution $P_{\mu^\ast}(n)$. (c)~Fitted $\Delta n$ as a function of $\tobs/\ta$ for several packing fractions (see legend). The dashed lines are exponential fits, from which we extract $\Delta n_\infty$. For the largest packing fraction $\phi = 0.57$, $\Delta n_\infty$ is estimated as a simple average. (d)~Extrapolated $\Delta n_{\infty}$ as a function of the packing fraction $\phi$. (e)~Scaled peak of the numerical susceptibility [symbols, see also (a)] and the predictions of Eq.~\eqref{eq:chimax:fss} (horizontal lines). (f)~Scaling of the chemical potential $\mu^\ast$ at coexistence. Dashed lines are linear fits to the data (symbols). (g)~Extrapolated $\mu^\ast_\infty$ as a function of (inverse) packing fraction. The solid line is a guide to the eye separating the low-LFS liquid from the high-LFS glass.}
  \label{fig:chi}
\end{figure*}

A bimodal shape of the probability distribution is not yet sufficient to claim a first-order transition. A hallmark of such a transition is the coexistence of both phases separated by an interface. In small systems (as studied here), the penalty associated with such an interface can be overcome by fluctuations and the system spontaneously transforms between the two phases. This is what we observe in Fig.~\ref{fig:coex}(a). As we approach the transition, numerical results are influenced by the (necessarily) finite box sizes that we can access in computers. A systematic study of these effects, however, has turned out to be a powerful tool to extract reliable results for phase boundaries and critical points~\cite{binder1981finite}. Here we perform such a finite-size scaling analysis varying the duration $\tobs$ of trajectories (equivalently, $K$) while keeping the number of particles $N$ fixed.

A central quantity to investigate phase transitions is the susceptibility $\chi(\mu)$ defined in Eq.~\eqref{eq:chi}, the response of the system with respect to an external change of parameters, here the chemical potential $\mu$. The susceptibility can be extracted from the fluctuations of the order parameter without an explicit perturbation and is shown in Fig.~\ref{fig:chi}(a) for $\phi=0.565$. The susceptibility $\chi(\mu)$ has the typical bell shape with a peak at $\mu^\ast$, which moves to smaller values as we increase $\tobs$. Concomitantly, the curve becomes narrower and its peak height $\chi^\ast\equiv\chi(\mu^\ast)$ grows. In the following, we use the phenomenological theory of first-order phase transitions of Binder~\cite{binder1984finite} to predict the asymptotic shape of the susceptibility in the limit of large $V$.

Following Ref.~\citenum{binder1984finite}, we posit that the reweighted distributions $P_\mu(n)$ can be approximated by 
\begin{equation}
  P_\mu(n) \approx (1 - p)P_\text{l}(n) + pP_\text{g}(n),
  \label{eq:P:gauss}
\end{equation}
where $P_\text{l,g}(n)$ are two Gaussian distributions, one for the liquid (l) and one for the glassy (g) phase. Each Gaussian is characterized by mean and variance, which correspond to the mean populations $n_\text{l,g}$ and the rescaled bulk susceptibilities $\chi_\text{l,g}/V$, respectively. We assume $\chi_\text{l,g}$ to be constant in the limit of large $V$. Both Gaussians in Eq.~\eqref{eq:P:gauss} are weighed by the probability $p=p(\mu;V)$ to find the system in the LFS-rich glass-like phase as a function of the chemical potential $\mu$. Figure~\ref{fig:chi}(b) shows $P_{\mu^\ast}(n)$ and the two fitted Gaussians (dashed lines) for an exemplary density $\phi=0.565$ and trajectory duration of $\tobs\simeq 96\ta$. The two Gaussians provide an accurate description of $P_{\mu^\ast}(n)$ close to the two peaks, but give a poor result in the vicinity of the minimum. This is due to a contribution of interfaces to the probability distribution, which is not accounted for in Eq.~\eqref{eq:P:gauss}. As we increase $\tobs$, this interfacial contribution becomes less important and we expect Eq.~\eqref{eq:P:gauss} to become accurate.

The superposition Eq.~\eqref{eq:P:gauss} implies that the susceptibility [Eq.~\eqref{eq:chi}] becomes
\begin{equation}
  \label{eq:chi:fss}
  \chi(\mu) = (1-p)\chi_l + p\chi_g + Vp(1-p)(\Delta n)^2
\end{equation}
with $\Delta n\equiv n_\text{g}-n_\text{l}$ and probability
\begin{equation}
  \label{eq:p:fss}
  p(\mu;V) = \left[1+\exp\left\{-V\Delta n(\mu-\mu^\ast)-2\frac{\chi_\text{g}-\chi_\text{l}}{V(\Delta n)^2}\right\}\right]^{-1},
\end{equation}
see Appendix~\ref{app:fss} for details. We fit Eq.~\eqref{eq:chi:fss} together with Eq.~\eqref{eq:p:fss} to the numerically determined susceptibilities, leaving $\Delta n$ as a free parameter and using $\chi_l$ and $\chi_g$ obtained from the fit of $P_{\mu^\ast}(n)$ with the superposition of the two Gaussians. The fits are shown as dashed lines in Fig.~\ref{fig:chi}(a). We obtain good agreement with the data, especially for large $V$ and close to coexistence ($\mu\simeq \mu^\ast$), where indeed the superposition of Gaussians [Eq.~\eqref{eq:P:gauss}] is expected to be accurate. In Fig.~\ref{fig:chi}(c), we plot the fit parameters $\Delta n$ as a function of $\tobs/\ta$ for several packing fractions. We observe that $\Delta n$ saturates at large $V$ to $\Delta n_\infty$. In Fig.~\ref{fig:chi}(d), $\Delta n_\infty$ is plotted and decreases as the packing fraction is increased.

From Eq.~\eqref{eq:chi:fss}, we also find that the peak height of the susceptibility $\chi^\ast$ approaches
\begin{equation}
  \frac{\chi^\ast}{V} \to \frac{(\Delta n_\infty)^2 }{4}
  \label{eq:chimax:fss}
\end{equation}
in the limit of large $V$. Employing the extracted $\Delta n_{\infty}$ [Fig.~\ref{fig:chi}(d)], we compare the prediction from Eq.~\eqref{eq:chimax:fss} to the observed peak susceptibilities as shown in Fig.~\ref{fig:chi}(e). Except for the highest packing fraction $\phi=0.57$, we observe that the data approaches the expected value as $\tobs\to\infty$, with linear corrections in $\tobs^{-1}$. The disagreement at the largest packing fraction $\phi=0.57$ can be attributed to the relatively short trajectories that we were able to access (compared to the relaxation time $\ta$).

From the peak of the numerical susceptibilities, we also extract the value $\mu^\ast$ of the chemical potential at coexistence, which is plotted in Fig.~\ref{fig:chi}(f). We find that it can be well fitted by the expression $\mu^\ast-\mu^\ast_\infty\propto(\ta/\tobs)$, which can be predicted from finite-size scaling theory (Appendix~\ref{app:fss}). In Fig.~\ref{fig:chi}(g), we plot $\mu^\ast_\infty$ as a function of the packing fraction.

\begin{figure}[t]
  \centering
  \includegraphics[width=\linewidth]{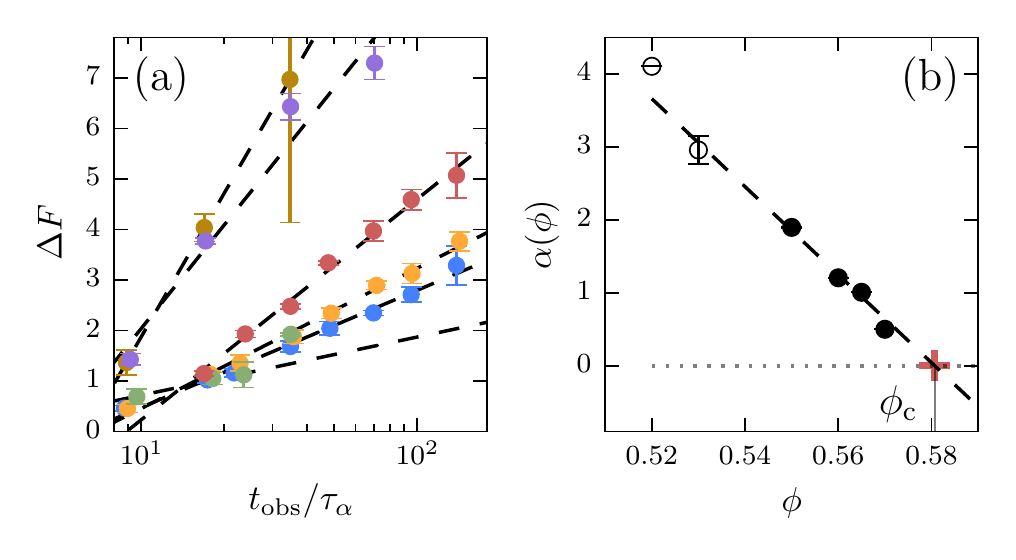}
  \caption{(a)~Effective barrier $\Delta F$ between the liquid peak and the minimum of $P_{\mu^\ast}(n)$ [see also Fig.~\ref{fig:chi}(b)]. Dashed lines are logarithmic fits (Eq.~\eqref{eq:alpha}) to the data (symbols). (b)~The slope $\alpha$ as a function of the packing fraction $\phi$, fitted (excluding the empty symbols) with a line (dashed), and its extrapolation to zero (red cross) at $\phi_\text{c}\simeq0.58$.}
  \label{fig:deltaF}
\end{figure}

Finally, we study the scaling of the barrier
\begin{equation}
  \Delta F = \ln P(n_\text{l}) - \ln P(n_\ast),
  \label{eq:deltaF}
\end{equation}
which measures the difference between the ``free-energy'' of the liquid-like phase (at $n_\text{l}$) and the local maximum between the two phases (at $n_\ast$) [cf. Fig.~\ref{fig:chi}(b)]. In analogy with conventional phase transitions, $\Delta F$ describes the effective barrier for a transition from the liquid to the glassy LFS-rich phase. As we elongate the trajectory, we observe a monotonous increase of $\Delta F$, which can be well described by the logarithmic functional form
\begin{equation}
  \Delta F = \alpha(\phi) \ln \left( \tobs / \ta \right),
  \label{eq:alpha}
\end{equation}
see Fig.~\ref{fig:deltaF}(a). We note that the increase of $\Delta F$ with $\tobs$ becomes weaker as the density is increased, \emph{i.e.}, $\alpha(\phi)$ decreases. Fig.~\ref{fig:deltaF}(b) shows that $\alpha(\phi)$ is decreasing linearly with $\phi$ and is extrapolated to reach zero at around $\phi_\text{c}\simeq 0.58$.
 
\section{Discussion}

\begin{figure}[b!]
  \centering
  \includegraphics[width=.8\linewidth]{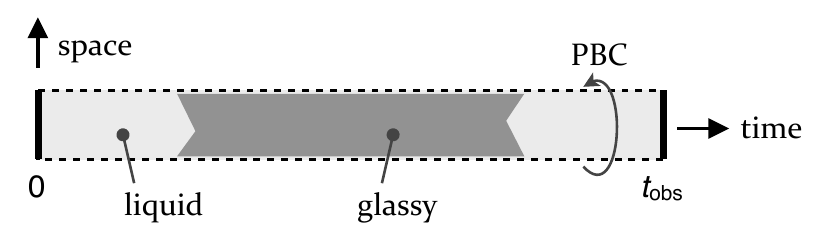}
  \caption{Sketch of the trajectory ``geometry''. We employ periodic boundary conditions (PBC) for the three spatial dimensions whereas time is bounded by ``walls'' at $t=0$ and $t=\tobs$ (thick lines). Indicated are two domains of liquid (light) and glassy (dark).}
  \label{fig:sketch}
\end{figure}

Following dynamic facilitation, structural relaxation is promoted through small and sparse excitations in the sea of immobile jammed particles~\cite{chandler2010dynamics}. Spatial coupling is weak (if not absent) so that these excitations form an ideal gas~\cite{keys2011excitations}. In contrast, the temporal coupling of excitations is strong (albeit still short-ranged), giving rise to a large temporal correlation length, \emph{i.e.}, structural relaxation time $\ta$. Excitations in spacetime are thus characterized by highly anisotropic interactions. Here we have studied the finite-size scaling along the dominant temporal direction, employing a structural order parameter on the premise that excitations carry a structural signature. For large $\tobs$ (keeping $N$ fixed), we observe within trajectories slow glassy domains with a large population of LFS coexisting with the supercooled liquid [Fig.~\ref{fig:coex}(a)]. Both regions are separated by an interface perpendicular to the time direction as sketched in Fig.~\ref{fig:sketch} in agreement with the notion of an effective interfacial tension minimizing the interfacial ``area''~\cite{katira2018solvation}. Note the different boundary conditions (Fig.~\ref{fig:sketch}): While we can employ periodic boundaries in space this is not possible for the temporal direction which can be viewed as confined by walls effectively repelling the glassy domain~\cite{garrahan2009first}. Nevertheless, we have demonstrated that fluctuations of the structural order parameter approach the asymptotic behavior predicted from two coexisting bulk phases [Fig.~\ref{fig:chi}]. Note that further increasing the trajectory length, one would expect a breaking of the single domain into multiple domains and ``domain breathing'' as observed in the Ising model~\cite{schmitz2014determination}.

Our results unambiguously demonstrate the existence of a discontinuous first-order phase transition in the ensemble of trajectories of polydisperse hard spheres between the normal liquid (low population of LFS and high mobility) and a non-equilibrium glassy phase (high population of LFS, the structural relaxation time exceeds the trajectory length). These results complement previous studies on two binary Lennard-Jones glass formers: Kob-Andersen~\cite{hedges2009dynamic,speck2012constrained,speck2012first,turci2018structural} and Wahnstr\"om~\cite{turci2018structural}.

\begin{figure}[t]
  \centering
  \includegraphics[width=0.5\textwidth]{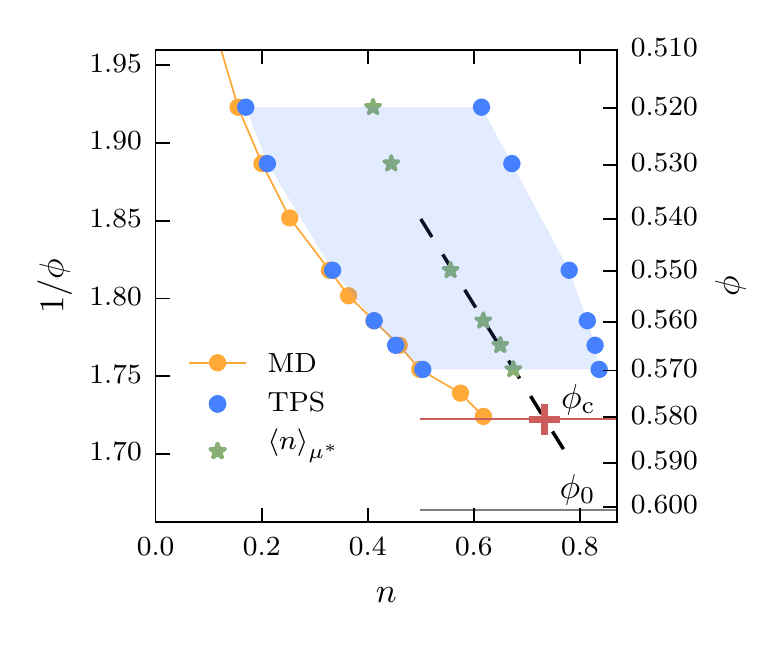}
  \caption{Coexistence region in the $(1/\phi,n)$ plane represented as the blue shaded area between the liquid ($n_\text{l}$) and glass ($n_\text{g}$) peaks from the probability distributions $P_{\mu^\ast}(n)$. Results are shown for biased TPS simulations (blue symbols) together with the population in the liquid phase obtained from straightforward EDMD simulations [orange symbols, also shown in Fig.~\ref{fig:system}(b)]. The dashed line is a linear fit of the ensemble-averaged population at coexistence $\mean{n}_{\mu^\ast}$ (green, star symbols), for $\phi\geqslant 0.55$. The grey horizontal line indicates the location of $\phi_0$, where the VFT expression for the relaxation time $\ta$ diverges. The red horizontal line indicates the location of $\phi_\text{c}$ where $\Delta F$ is expected to become independent of $\tobs$.}
  \label{fig:phasedia}
\end{figure}

In Fig.~\ref{fig:chi}(g), we show the phase diagram for the two intensive variables $1/\phi$ (the control parameter) and $\mu$ (the external field), whereby along the solid line the order parameter $\mean{n}_{\mu}$ changes discontinuously. Switching to the extensive variable $n$, the line widens into the coexistence region, which is shown in Fig.~\ref{fig:phasedia}. Coexisting populations are extracted from the liquid and glassy peaks of $P_{\mu^\ast}(n)$ at $\mu=\mu^\ast$ (using the ensembles with the largest $\tobs$). We can only construct a section of the full phase diagram for moderate densities. Beyond $\phi=0.57$, equilibrium sampling of trajectories is computationally too expensive with our simulation scheme due to the steep increase of the structural relaxation time. Going to lower densities, even though the relaxation time is short, the structural-dynamical transition moves away from $\mu=0$ [cf. Fig.~\ref{fig:chi}(g)]. Sampling of trajectories then requires a prohibitively large number of TPS steps to reach equilibration. As a consequence, it is not yet clear how much further the coexistence region extends.

In the Ising model, the coexistence region is terminated by a critical point. Indeed, in Fig.~\ref{fig:phasedia} we observe that the coexistence region narrows as we go to higher packing fractions $\phi$. Moreover, the extrapolation of the slope $\alpha(\phi)$ is expected to reach zero at $\phi_c\simeq 0.58$ [see Fig.~\ref{fig:deltaF}(b)], implying that the free-energy barrier $\Delta F$ becomes independent of the finite size of the system. This is indeed expected for a critical point. We thus have some numerical evidence for a critical point beyond which the two phases would become indistinguishable at least structurally. Such a critical point has been proposed previously also for the Kob-Andersen binary mixture~\cite{turci2017nonequilibrium}, and thus might be a general feature of glass formers.

Our estimated $\phi_c\simeq 0.58$ is within the region of packing fractions that is accessible to direct numerical investigations using molecular dynamics or Monte Carlo simulations. So far, no transition from an LFS-poor to an LFS-rich phase has been reported. There are several possible reasons. First, the critical point might be at $\mu_\text{c}>0$ and thus not accessible by unbiased dynamics. In Fig.~\ref{fig:chi}(g), we see that $\mu^\ast_\infty$ seems to approach zero but we cannot rule out that $\mu_\text{c}$ is non-zero (albeit small). Moreover, while we probe the regime of sufficiently long trajectories, the number $N=100$ of particles is small. As a result, the critical point might depend on $N$ and consequently move to higher densities as $N$ is increased. That swap simulations are able to equilibrate a hard-sphere fluid well beyond $\phi_\text{c}$ is compatible with our results as it demonstrates the absence of a thermodynamic cause of dynamic arrest, while here it is explicitly attributed to the local dynamics.


\section{Conclusions}

By performing extensive numerical simulations of a hard-sphere model glass former, we have investigated the finite-size scaling of a dynamical order-disorder transition in the space of trajectories. We observe two phases characterized by trajectories with different dynamics and different population of the model's LFS, the defective icosahedron. The first phase has a low population of the LFS, its relaxation time matches $\ta$, and it is indistinguishable from the equilibrium liquid. The second phase possesses a higher population of the LFS and has a relaxation time which is typically longer than the duration of the trajectories we can access, making it a non-equilibrium glass.

We have chosen the population $n$ of the LFS as an order parameter to distinguish the two phases, and computed its probability distribution using a computational scheme based on TPS. The choice of $n$ is motivated by the strong correlation with the relaxation time $\ta$ [see Fig.~\ref{fig:system}], which hints at a connection between the glassy slowdown and the emergence of local structures. The observed phenomenology agrees with a first-order phase transition due to short-ranged interactions. The phase diagram is reminiscent of the one observed in Ref.~\citenum{turci2017nonequilibrium} for the Kob-Andersen binary mixture, where the existence of two non-equilibrium critical points at high and low temperatures was suggested. By extrapolation to packing fractions beyond the computationally accessible range, we find indications of the presence of a critical point at high packing fractions. The existence of a critical point, observed also in softened kinetically constrained models~\cite{elmatad2010finite}, could thus prove to be a general feature -- at least of a certain class -- of glass formers. While hard spheres exhibit the necessary strong correlation between local structure and global dynamics, other glass formers might not: \emph{e.g.} polymeric glass formers, where the identification of an LFS is challenging, and systems where known candidates for LFS are only weakly correlated with the dynamical slowdown~\cite{hocky2014correlation}.


\begin{acknowledgements}
  We thank C.P. Royall and F. Turci for helpful discussions. Early stages of this work have been funded through the DFG School of Excellence ``Materials Science in Mainz'' (GSC 266). We gratefully acknowledge computing time on the supercomputer MOGON II at Johannes Gutenberg University Mainz (hpc.uni-mainz.de).
\end{acknowledgements}


\appendix

\section{Single particle dynamics}
\label{app:jumps}

\begin{figure}[t]
  \centering
  \includegraphics{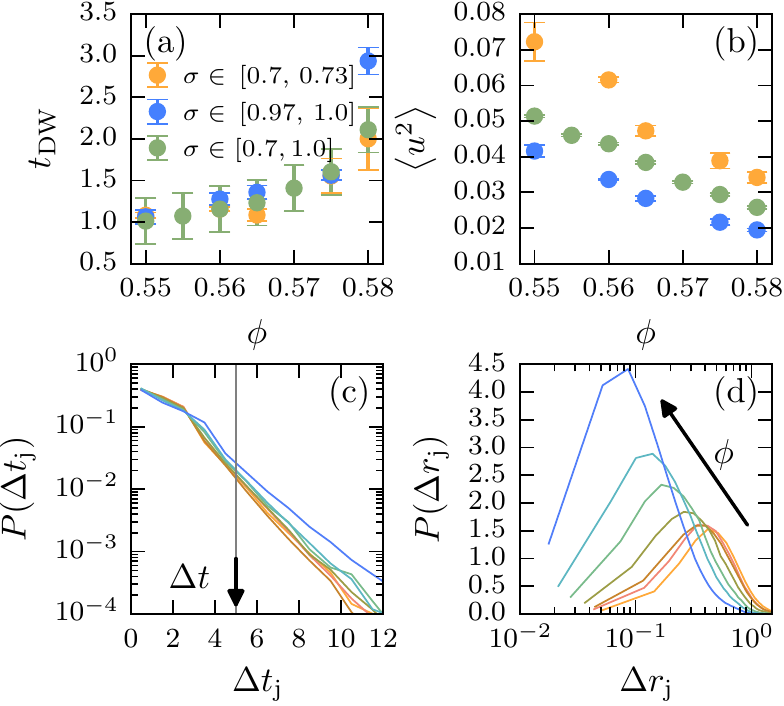}
  \caption{Single particle dynamics. (a)~Time $t_{\mathrm{DW}}$ and (b)~Debye-Waller factor $\mean{u^2}$ for all particles (green symbols), $10\%$ largest particles ($\sigma \in \left[ 0.97, 1.0 \right]$, blue symbols), and $10\%$ smallest particles ($\sigma \in \left[ 0.7, 0.73 \right]$, orange symbols). (c)~Distribution of jump durations $\Delta t_{\text{j}}$ and (d)~jump lengths $\Delta r_{\text{j}}$ for $\phi \in \left[0.55, 0.58\right]$ (from orange to blue).}
  \label{fig:DW}
\end{figure}

In the high-density regime, particles in a liquid are trapped by their nearest neighbors, a phenomenon referred to as caged dynamics. When space is made available, particles can escape by ``jumping'' to a different cage.
We use the time extent of these jumps to define the elementary unit of time  $\Delta t$ for coarse-graining our trajectories. Following Ref.~\citenum{pastore2014cage}, we first estimate the average size of the cages using the Debye-Waller factor
\begin{equation}
  \mean{u^2} = \min_{t} \left[ \frac{ \partial \ln{\mathrm{MSD}(t)} }{ \partial \ln{t} } \right],
  \label{eq:debye-waller}
\end{equation}
where $\text{MSD}$ is the mean-squared displacement.
The time $t_{\text{DW}}$ which realizes the minimum in Eq.~\eqref{eq:debye-waller} exists only in the supercooled/dense regime, where the $\text{MSD}$ develops the typical plateau. We calculate $t_{\text{DW}}$ and $\mean{u^2}$ in the range $\phi \in \left[0.55, 0.58\right]$ for the $10 \%$ smallest, the $10 \%$ largest and for all particles. We find that $t_{\text{DW}}$ increases with $\phi$ and it is largely independent on the size of the particle, except for $\phi = 0.58$ where the largest particles show an increase of a factor of 1.5 with respect to the average [Fig.~\ref{fig:DW}(a)]. The Debye-Waller factor decreases with $\phi$ and with the size of the particle [Fig.~\ref{fig:DW}(b)].

We define a particle to be in the jumping state at time $t$ if its mean-squared displacement, calculated over the interval $\left[ t - 1, t + 1 \right]$, exceeds $\mean{u^2}$. The time and space extents of a jump are referred to as $\Delta t_{\text{j}}$ and $\Delta r_{\text{j}}$ respectively, and we compute their distributions [Fig.~\ref{fig:DW}(c,d)]. We find that the distribution of $\Delta t_{\text{j}}$ decays exponentially and it is largely independent on $\phi$, that $\Delta r_{\text{j}}$ is typically smaller than the size of the particles and depends sensitively on $\phi$. In order to include the vast majority of the jumps, we choose $\Delta t = 5$, which corresponds to the time at which $P(\Delta t_{\text{j}})$ has decayed by approximately one order of magnitude from its maximum value. The precise choice of $\Delta t$ does not influence qualitatively our results, but it is important to choose a value that is not too small ($\Delta t < 1$), in order not to oversample the trajectory in time.

\section{Finite-size scaling theory}
\label{app:fss}

To obtain an expression for the probability $p(\mu;V)$, we consider the restricted partition functions
\begin{equation}
  \label{eq:app:Z}
  Z_i(\mu;V) = \IInt{n}{n\in i}{} P(n;V)e^{\mu Vn}
\end{equation}
for liquid and glass through integrating over the intervals $i$ indicated as shaded areas in Fig.~\ref{fig:coex}(b), with which
\begin{equation}
  p(\mu;V) = \frac{Z_\text{g}(\mu;V)}{Z_\text{g}(\mu;V)+Z_\text{l}(\mu;V)}.
\end{equation}
Expanding close to coexistence leads to
\begin{equation}
  \ln Z_i(\mu;V) \approx \ln Z_i(\mu^\ast;V) + Vn_i(\mu-\mu^\ast) + \dots
\end{equation}
To obtain an expression at $\mu^\ast$ (which itself depends on system size), we use
\begin{equation}
  p(\mu^\ast) = \frac{1}{2} + \frac{\chi_\text{g}-\chi_\text{l}}{2V(\Delta n)^2}
\end{equation}
obtained from solving $\pd{\chi}{p}=0$ for the susceptibility given in Eq.~\eqref{eq:chi:fss}. Putting everything together, we thus arrive at the expression given in Eq.~\eqref{eq:p:fss}.

To find the dependence of $\mu^\ast(V)$, we equate the partition functions Eq.~\eqref{eq:app:Z} after expanding around $\mu^\ast_\infty=\mu^\ast(V\to\infty)$,
\begin{equation}
  -\gam_\infty N + V\Delta n_\infty(\mu^\ast-\mu^\ast_\infty) = 0,
\end{equation}
where now we have to include an effective interfacial tension $\gam_\infty$ and at coexistence $Z_\text{g}(\mu^\ast_\infty)=Z_\text{l}(\mu^\ast_\infty)$. Hence, close to coexistence we find $\mu^\ast-\mu^\ast_\infty\propto(\tobs/\ta)^{-1}$.


%

\end{document}